\documentclass{PoS}
\usepackage{amsmath}

\def\Code#1{\ensuremath{\texttt{#1}}}
\def\Var#1{\ensuremath{\mathit{#1}}}
\def\Vn{\Var{n}}
\def\Vf{\Var{f}}
\def\ie{i.e.\ }
\def\eg{e.g.\ }
\def\lbrac{\symbol{123}}
\def\rbrac{\symbol{125}}
\def\Brac#1{\lbrac#1\rbrac}
\def\unity{{\rm 1\mskip-4.25mu l}}

\title{FormCalc 6}

\ShortTitle{FormCalc 6}

\author{\speaker{Thomas Hahn} \\
        Max-Planck-Institut f\"ur Physik \\
	F\"ohringer Ring 6 \\
	D--80805 M\"unchen \\
        E-mail: \email{hahn@feynarts.de}}

\abstract{The new features and improvements in FormCalc Version 6 as 
well as some recent additions in FeynArts for easier diagram selection 
are reported.}

\FullConference{XII Advanced Computing and Analysis Techniques in Physics Research\\
		November 3--7 2008\\
		Erice, Italy}

\begin{document}

\section{Introduction}

FeynArts \cite{FeynArts} and FormCalc \cite{FormCalc} generate and 
calculate Feynman diagrams to the extent that one can easily obtain the 
analytic expression of an amplitude at up to one-loop order and create a 
Fortran code for the numerical evaluation of the squared matrix element 
almost fully automatically.

This note presents some recent improvements and additions to both 
programs.  Firstly, it elaborates on diagram filters in FeynArts, for 
which several ancillary functions have been added.  Secondly, the 
following new features in FormCalc 6 are described:
\begin{itemize}
\item Alternate-channel communication between FORM and Mathematica.

\item Code-generation for CutTools.

\item Improvements in Dirac chains (4D) for analytical purposes.

\item Functions to re-use previously introduced abbreviations
      and subexpressions in a new session.

\item The \Code{Keep} function to resume interrupted calculations.

\item New \Code{Abbreviate} mode.

\item Options to fine-tune code generation.
\end{itemize}


\section{Diagram filters in FeynArts}

The \Code{DiagramSelect} function in FeynArts performs on Feynman 
diagrams (\ie the output of \Code{InsertFields}, denoted by `\Code{ins}' 
in the following examples) what \Code{Select} does to ordinary 
Mathematica expressions: it retains those for which a given test 
function returns \Code{True}.

In the case of \Code{DiagramSelect}, the test function receives three 
arguments:
\begin{enumerate}
\item the insertions, of the form 
\Code{Graph[$h$][Field[1]$\,\to f_1$,\,Field[2]$\,\to f_2$,\,$\ldots$]},

\item the topology belonging to these insertions,

\item the header of the topology list to which the topology belongs.
\end{enumerate}
The \Code{DiagramSelect} function is part of FeynArts already for quite 
a while and the information it provides to the test function allows the 
skilled programmer a very detailed selection.  Recently, a number of 
functions have been added with which the construction of sophisticated 
test functions is straightforward.  Unless otherwise stated, the 
functions introduced in the following subsections take the three 
arguments detailed above.


\subsection{\Code{LoopFields}}

\Code{LoopFields} returns a list of fields running in the loop(s) of the 
given diagram.  It might be used as in
\begin{verbatim}
DiagramSelect[ins, FreeQ[LoopFields[##], V[1]]&]
\end{verbatim}
to omit all diagrams where a photon runs in the loop.  Or, to select 
only diagrams of $\mathcal{O}(\alpha_s)$ or higher with respect to
tree-level, one could use:
\begin{verbatim}
CountQCDFields[f_] :=
  Count[f, _. (F[3|4, ___] | V[5, ___] | U[5, ___])];
DiagramSelect[ins, CountQCDFields[LoopFields[##]] > 2 &]
\end{verbatim}


\subsection{\Code{WFCorrectionFields}, \Code{WFCorrectionCTFields}}

\Code{WFCorrection(CT)Fields} returns the fields external to
wave-function corrections, \ie self-energies on external legs (or their
counter-terms in the case of \Code{WFCorrectionCTFields}).  For example, 
if the diagram contains a self-energy insertion 
$\Code{S[1]}\to\Code{S[3]}$ on an external leg, the result of 
\Code{WFCorrectionFields} is \Code{\Brac{S[1], S[3]}}.  If the diagram 
contains no wave-function correction, the list is empty.

This filter is typically used to eliminate wave-function corrections 
with identical external legs, \ie remove corrections of the type $a\to 
a$ but keep $a\to b$ (for, the former contain a denominator $1/(p^2 - 
m_a^2)$ which, if naively inserted, evaluates to $1/(m_a^2 - m_a^2)$ for 
an on-shell $p$).  This can be done with a construction like
\begin{verbatim}
DiagramSelect[ins, UnsameQ@@ WFCorrectionFields[##] &]
\end{verbatim}   


\subsection{\Code{Vertices}, \Code{FieldPoints}}

\Code{Vertices} returns the vertices contained in a topology, not
counting the endpoints of the external legs (even though they are
internally represented as \Code{Vertex[1][\Vn]}).  Note that this is a
purely topological list of vertices, with no field information.

\Code{FieldPoints} returns the field content for each vertex of a 
topology, \ie a list of objects of the form 
\Code{FieldPoint[\Var{cto}][\Var{fields}]}, where \Var{cto} is the 
counter-term order.

These functions are most efficiently used together with the following.


\subsection{\Code{FieldMatchQ}, \Code{FieldMemberQ}, 
\Code{FieldPointMatchQ}, \Code{FieldPointMemberQ}}

\Code{FieldMatchQ[\Vf,\,\Var{patt}]} returns \Code{True} if the field 
\Vf\ matches the pattern \Var{patt} and \Code{False} otherwise.  It 
works like \Code{MatchQ} but takes into account field levels, \eg 
\Code{F[1]} matches \Code{F}.

\Code{FieldMemberQ[\Var{flist},\,\Var{patt}]} returns \Code{True} if an 
element of \Var{flist} matches \Var{patt} in the sense that 
\Code{FieldMatchQ} returns \Code{True}.

\Code{FieldPointMatchQ[\Var{fp},\,\Var{patt}]} returns \Code{True} if 
the field point \Var{fp} matches the pattern \Var{patt} and \Code{False} 
otherwise.  The matching takes into account field levels, \eg 
\Code{F[1]} matches \Code{F}.

\Code{FieldPointMemberQ[\Var{fplist},\,\Var{patt}]} returns \Code{True} 
if an element of \Var{fplist} matches \Var{patt} in the sense that 
\Code{FieldPointMatchQ} returns \Code{True}.

With these functions, it is quite simple to select diagrams with a 
particular field point, \eg
\begin{verbatim}
DiagramSelect[ins, FieldPointMemberQ[FieldPoints[##],
  FieldPoint[V[1], F[2], -F[2]]]&]
\end{verbatim}
selects only diagrams with a photon--lepton--lepton coupling.


\section{Alternate Link between FORM and Mathematica}

FORM is renowned for being able to handle very large expressions.  To 
yield (pre-)simplified expressions, however, terms have to be wrapped 
in functions, to avoid immediate expansion.  The number of terms in a 
function is rather severely limited in FORM: on 32-bit systems to 32568.

FormCalc is thus in a dilemma: it has become much more sophisticated in 
simplifying amplitudes in recent versions and partly because of this, 
users now compute larger amplitudes.  This is the reason that many users 
have recently reported (unorthodox) `overflow' messages from FORM.

The solution introduced in Version 6 is to send the pre-simplified 
generic amplitude to Mathematica intermediately for introducing 
abbreviations.  To enable this out-of-band communication, the ReadForm 
executable (which manages the traffic between FORM and Mathematica) 
threads off a pipe handler which communicates with FORM via its external 
channels \cite{extform}.

The result is a significant reduction in size of the intermediate 
expressions, shown here for the u-channel part of the tree-level 
amplitude of $uu\to gg$.  The pre-simplified generic expression passed 
from FORM to Mathematica is in this case:

\begin{footnotesize}
\begin{verbatim}
Den[U,MU2]*(
-8*SUNSum[Col5,3]*SUNT[Glu3,Col5,Col2]*SUNT[Glu4,Col1,Col5]*mul[Alfas*Pi]*
  abb[fme[WeylChain[DottedSpinor[k1,MU,-1],6,Spinor[k2,MU,1]]]*ec3.ec4
      -1/2*fme[WeylChain[DottedSpinor[k1,MU,-1],6,ec3,ec4,Spinor[k2,MU,1]]]
      +fme[WeylChain[DottedSpinor[k1,MU,-1],7,Spinor[k2,MU,1]]]*ec3.ec4
      -1/2*fme[WeylChain[DottedSpinor[k1,MU,-1],7,ec3,ec4,Spinor[k2,MU,1]]]]*MU
-4*SUNSum[Col5,3]*SUNT[Glu3,Col5,Col2]*SUNT[Glu4,Col1,Col5]*mul[Alfas*Pi]*
  abb[fme[WeylChain[DottedSpinor[k1,MU,-1],6,ec3,ec4,k3,Spinor[k2,MU,1]]]
      -2*fme[WeylChain[DottedSpinor[k1,MU,-1],6,ec4,Spinor[k2,MU,1]]]*ec3.k2
      -2*fme[WeylChain[DottedSpinor[k1,MU,-1],6,k3,Spinor[k2,MU,1]]]*ec3.ec4
      +fme[WeylChain[DottedSpinor[k1,MU,-1],7,ec3,ec4,k3,Spinor[k2,MU,1]]]
      -2*fme[WeylChain[DottedSpinor[k1,MU,-1],7,ec4,Spinor[k2,MU,1]]]*ec3.k2
      -2*fme[WeylChain[DottedSpinor[k1,MU,-1],7,k3,Spinor[k2,MU,1]]]*ec3.ec4]
+8*SUNSum[Col5,3]*SUNT[Glu3,Col5,Col2]*SUNT[Glu4,Col1,Col5]*mul[Alfas*MU*Pi]*   
  abb[fme[WeylChain[DottedSpinor[k1,MU,-1],6,Spinor[k2,MU,1]]]*ec3.ec4
      -1/2*fme[WeylChain[DottedSpinor[k1,MU,-1],6,ec3,ec4,Spinor[k2,MU,1]]]
      +fme[WeylChain[DottedSpinor[k1,MU,-1],7,Spinor[k2,MU,1]]]*ec3.ec4
      -1/2*fme[WeylChain[DottedSpinor[k1,MU,-1],7,ec3,ec4,Spinor[k2,MU,1]]]] )
\end{verbatim}
\end{footnotesize}
and the version coming back from Mathematica is:
\begin{footnotesize}
\begin{verbatim}
-4*Den(U,MU2)*SUNSum(Col5,3)*SUNT(Glu3,Col5,Col2)*SUNT(Glu4,Col1,Col5)*
  AbbSum5*Alfas*Pi 
\end{verbatim}
\end{footnotesize}


\section{CutTools}

The CutTools package evaluates one-loop integrals via the 
cutting-technique-inspired Ossola, Papadopoulos, Pittau (OPP) method 
\cite{CutTools}.

FormCalc 6 can generate code for linking with CutTools.  This is 
governed by the \Code{CutTools} option of \Code{CalcFeynAmp}, which has 
three values: \Code{False} selects LoopTools functions \cite{FormCalc}, 
\ie traditional Passarino--Veltman tensor reduction, \Code{True} selects 
CutTools functions, but with the rational terms worked out by FormCalc 
analytically, and \Code{Rational} selects CutTools functions with the 
rational terms computed by CutTools.

The CutTools functions emitted by FormCalc form a set \Code{Acut}, 
\Code{Bcut}, etc., analogous to the \Code{A0i}, \Code{B0i}, etc.\ 
LoopTools functions, \ie the Passarino--Veltman tensor coefficient 
functions.  From the technical point of view, the CutTools functions 
require the numerator as a function of the integration momentum $q$ 
which it can sample to solve the cutting equations.  The correspondence 
is, for example,
\begin{align*}
\int\frac{(\mathrm{i}\pi^2)^{-1}\ \mathrm{d}^4 q\ q_\mu q_\nu}
         {(q^2 - m_1^2) ((q - p)^2 - m_2^2)}
&= g_{\mu\nu}\,\Code{B0i}(\Code{bb00}, p^2, m_1^2, m_2^2) +
   p_\mu p_\nu\,\Code{B0i}(\Code{bb11}, p^2, m_1^2, m_2^2) \\
&= \Code{Bcut}(2, \Code{num1}, \Code{num2}, p, m_1^2, m_2^2)\,.
\end{align*}
The first line is the classical Passarino--Veltman decomposition, 
formulated in terms of the LoopTools function \Code{B0i}.  The second 
line is the two-point CutTools function \Code{Bcut}.

The main novelty here are the two numerator functions $\Code{num1} = 
q_\mu q_\nu$ and $\Code{num2} = 0$, the latter of which is the 
coefficient of $D - 4$ and is sampled only if CutTools is used to 
compute also the rational terms.  Although not evident from this 
example, the numerators are in general scalar functions, as the $q_\mu$ 
are contracted with other vectors, are part of a fermion chain, or 
similar.  The 2 describes the maximum tensor rank ($q_\mu q_\nu$ = 2 
instances of $q$).  Note further that the momentum $p$ is passed
directly (not squared), as the routine may have to compute dot products 
etc.

The final link with CutTools, including proper detection of the package 
in \Code{configure}, is not yet complete but will be available shortly. 
For this reason, no performance reports can yet be given.  Regardless of 
any performance questions, however, the CutTools option in itself is 
definitely important as an independent way of checking LoopTools 
results.


\section{Dirac Chains in 4D}

As numerical calculations involving external fermions are mostly done
using Weyl-spinor chains, there has been a paradigm shift for Dirac 
chains to make them better suited for analytical purposes, \eg the 
extraction of Wilson coefficients.

Already in Version 5, with improvements in Version 6, Fierz methods have 
been implemented for Dirac chains, thus allowing the user to force the 
Dirac chains into almost any desired order, via the \Code{FermionOrder} 
option of \Code{CalcFeynAmp}.

Version 6 adds the \Code{Colour} method to the \Code{FermionOrder} 
option, which brings the spinors into the same order as the external 
colour indices.  This is not entirely trivial internally because colour 
is treated as an insertion, \ie is not present during the simplification 
of the generic amplitude.

Also new in Version 6 is the \Code{Antisymmetrize} option of 
\Code{CalcFeynAmp} which antisymmetrizes Dirac chains.  Such chains can 
be recognized by a negative chirality identifier, \eg 
$\Code{DiracChain[-6,\,$\ldots$]}$ instead of 
$\Code{DiracChain[6,\,$\ldots$]}$.  More precisely,
$$
\Code{DiracChain[$-\lambda$,\,$\mu_1$,$\ldots$,$\mu_n$]} =
\frac 1{n!}\sum_{\pi}\sigma_\pi
\omega_\lambda\gamma_{\pi(\mu_1)}\cdots\gamma_{\pi(\mu_n)}
$$
where the sum extends over all permutations $\pi$ of
$\mu_1,\ldots,\mu_n$, $\sigma_\pi$ is the sign of the permutation, and
$\omega_1 = \unity$, $\omega_5 = \gamma_5$, $\omega_{6,7} = \frac 12
(\unity \pm \gamma_5)$.  This notation includes as a special case
$\sigma_{\mu\nu} = \frac 12 (\gamma_\mu\gamma_\nu -
\gamma_\nu\gamma_\mu) = \Code{DiracChain[-1,\,$\mu$,\,$\nu$]}$.


\section{Re-using Abbreviations}

Abbreviations are a fundamental concept in FormCalc to reduce the size 
of expressions.  Abbreviations were so far restricted to one FormCalc 
session, however, \eg one could not save intermediate results involving 
abbreviations and resume computation in a new session.  FormCalc 6 adds 
two functions to `register' abbreviations and subexpressions from an 
earlier session.

\Code{RegisterAbbr} registers a list of abbreviations, \eg the output of
\Code{Abbr[]} in a previous session, such that future invocations of
\Code{CalcFeynAmp} will make use of them.  Note that abbreviations
introduced for different processes are in general not compatible.

\Code{RegisterSubexpr} registers a list of subexpressions, \eg the
output of \Code{Subexpr[]} in a previous session, such that future
invocations of \Code{Abbreviate} will make use of them.


\section{\Code{Keep}}

The \Code{Keep} function is an application of \Code{RegisterAbbr} and 
\Code{RegisterSubexpr} and is helpful in long-running calculations to 
store intermediate expressions, such that the calculation can be resumed 
after a crash.  As a side effect, the intermediate results can be 
inspected easily, even while a batch job is in progress.

\Code{Keep} has two basic arguments: a file (path and name) and an
expression.  If the file exists, it is loaded.  If not, the expression
is evaluated and the results stored in the file, thus creating a
checkpoint.  If the calculation crashes, it suffices to restart the very
same program, which will then load all parts of the calculation that
have been completed and resume at the point it left off.

\Code{Keep[\Var{expr},\,\Var{name},\,\Var{path}]} loads 
``\Code{\Var{path}/\Var{name}.m}'' if it exists, otherwise evaluates 
\Var{expr} and stores the result (together with the output of 
\Code{Abbr[]} and \Code{Subexpr[]}) in that file.  \Var{path} is 
optional and defaults to \Code{\$KeepDir} (= \Code{"keep"} by default).

\Code{Keep[\Var{lhs} = \Var{rhs}]} is the same as \Code{\Var{lhs} = 
Keep[\Var{rhs},\,"\Var{lhs}"]}.  This second syntax makes adding 
\Code{Keep} functionality to existing programs very simple.  For 
example, a statement like
\begin{verbatim}
amps = CalcFeynAmp[...]
\end{verbatim}
simply becomes
\begin{verbatim}
Keep[amps = CalcFeynAmp[...]]
\end{verbatim}
Since the file name is derived from the assigned-to identifier, this 
logic fails to work if symbols are being re-assigned, \ie appear more 
than once on the left-hand side, as in
\begin{verbatim}
Keep[amps = CalcFeynAmp[virt]]
Keep[amps = Join[amps, CalcFeynAmp[counter]]
\end{verbatim}
Due to the first \Code{Keep} statement, the second will always find the
file \Code{keep/amps.m} and never execute the computation of the counter
terms.

There are other ways to confuse the system, which usually require some 
deliberation, however: mixing intermediate results from different 
calculations, changing flags out of sync with the intermediate results, 
etc.  In case of doubt, \ie if results seem suspicious, remove all 
intermediate files and re-do the calculation from scratch.


\section{New \Code{Abbreviate} mode}

The \Code{Abbreviate} function, first introduced in Version 5, lets the 
user introduce abbreviations for arbitrary expressions.  The main deal 
here is the optimization performed during Fortran-code generation, where 
the abbreviations are grouped into categories, such that each one is 
evaluated as few times as possible -- \eg for MSSM calculations this 
routinely leads to speed-ups of $\approx$ 3.

The only invocation so far was 
\Code{Abbreviate[\Var{expr},\,\Var{lev}]}, where the integer \Var{lev} 
was the level inside \Var{expr} below which abbreviations were 
introduced and hence controlled how much of \Var{expr} was `abbreviated 
away.'  This has been complemented by the 
\Code{Abbreviate[\Var{expr},\,\Var{patt}]} mode, which introduces 
abbreviations for all things free of the pattern \Var{patt}.  Aside from 
performance issues, this is useful to get a picture of the structure of 
an expression.  For example,
\begin{verbatim}
Abbreviate[a + b + c + (d + e) x, x, MinLeafCount -> 0]
\end{verbatim}
gives \Code{Sub2 + Sub1\:x}, thus indicating that the original 
expression is linear in \Code{x}.  (Note that the \Code{MinLeafCount} 
option here is necessary because by default no abbreviations would be 
introduced for such simple subexpressions.)  As before, \Code{Subexpr[]} 
recovers the subexpressions.


\section{Varia}

\Code{WriteSquaredME} and \Code{WriteRenConst} have new options to 
fine-tune Fortran-code generation.  This is useful if one needs the 
generated code for purposes other than compilation with FormCalc's own 
driver programs.  The \Code{FileIncludes} and \Code{SubroutineIncludes} 
options respectively specify per-file and per-subroutine 
\Code{\#include} statements or similar declaration code, and 
\Code{FileHeader} gives the header written to each generated Fortran 
file.

The ReadForm MathLink executable has been made much more stable with 
respect to user aborts (Ctrl-C), \ie ReadForm zombie processes eating up 
CPU time will hopefully be a relict of the past now.


\section{Summary}

The latest FeynArts release (www.feynarts.de) features a number of 
functions which aid the user in constructing selection functions for 
\Code{DiagramSelect}.

The new FormCalc Version 6 (www.feynarts.de/formcalc) has many new and 
improved features.  The most important change is one the user will 
scarcely notice: the exchange of expressions with Mathematica halfway 
through the evaluation in FORM, which removes a bottleneck for largish 
amplitudes.  The support for CutTools is complete as far as the 
algebraic generation of amplitudes; the remaining few (mostly technical) 
details regarding linking etc.\ will be finalized soon.



\begin{thebibliography}{99}

\bibitem{FeynArts}
T.~Hahn, \textsl{Comp.\ Phys.\ Commun.} \textbf{140} (2001) 418
[hep-ph/0012260].

\bibitem{FormCalc}
T.~Hahn, M.~P\'erez-Victoria, \textsl{Comp.\ Phys.\ Commun.}
\textbf{118} (1999) 153 [hep-ph/9807565].

\bibitem{extform}
M.~Tentyukov, J.~Vermaseren, \textsl{Comp.\ Phys.\ Commun.} 
\textbf{176} (2007) 385 [cs/0604052].

\bibitem{CutTools}
G.~Ossola, C.~Papadopoulos, R.~Pittau 
\textsl{JHEP} \textbf{0803} (2008) 042 [arXiv:0711.3596].

\end{thebibliography}
\end{document}